\documentstyle[emulateapj]{article}

\def\simlt{\lower.5ex\hbox{$\; \buildrel < \over \sim \;$}}
\def\simgt{\lower.5ex\hbox{$\; \buildrel > \over \sim \;$}}
\def\gsim{\;\rlap{\lower 2.5pt
\hbox{$\sim$}}\raise 1.5pt\hbox{$>$}\;}
\def\lsim{\;\rlap{\lower 2.5pt
   \hbox{$\sim$}}\raise 1.5pt\hbox{$<$}\;}
\def\msun{{\rm\,M_\odot}}

\def\spose#1{\hbox to 0pt{#1\hss}}
\def\lta{\mathrel{\spose{\lower 3pt\hbox{$\mathchar''218$}}
     \raise 2.0pt\hbox{$\mathchar''13C$}}}
\def\gta{\mathrel{\spose{\lower 3pt\hbox{$\mathchar''218$}}
     \raise 2.0pt\hbox{$\mathchar''13E$}}}
\newcommand{\beq}{\begin{equation}}
\newcommand{\eeq}{\end{equation}}
\newcommand{\rem}[1]{{ }}


\lefthead{MEDVEDEV \& MENOU}
\righthead{HOT ACCRETION IN QUIESCENT DWARF NOVAE}

\begin{document}

\title{Hot Accretion onto White Dwarfs in Quiescent Dwarf Novae}

\author{Mikhail V. Medvedev\altaffilmark{a} \& Kristen Menou\altaffilmark{b,1} }

\affil{$^{a}$Canadian Institute for Theoretical Astrophysics, University of Toronto, Toronto, ON, M5S 3H8, Canada; medevedev@cita.utoronto.ca}
\affil{$^{b}$Princeton University, Department of Astrophysical Sciences, Princeton, NJ 08544; kristen@astro.princeton.edu}

\altaffiltext{1}{Chandra Fellow}

\authoremail{medevedev@cita.utoronto.ca}
\authoremail{kristen@astro.princeton.edu}

\vspace{\baselineskip}

\begin{abstract}

We present dynamically consistent solutions for hot accretion onto
unmagnetized, rotating white dwarfs (WDs) in five quiescent dwarf
novae. The measured WD rotation rates (and other system parameters) in
RX And, SS Cyg, U Gem, VW Hyi and WZ Sge imply spindown of the WD by
an extended hot flow emitting most of its X-rays in the vicinity of
the stellar surface.  In general, energy advection is absent and the
flow is stable to convection and hydrodynamical outflows. In rapidly
rotating systems, the X-ray luminosity provides only an upper limit on
the quiescent accretion rate because of substantial stellar spindown
luminosity. We suggest that the presence of hot flows in quiescent
dwarf novae may limit the long-term WD rotation rates to significantly
sub-Keplerian values.

\end{abstract}



\section{Introduction}

Dwarf novae (DN) are accreting binary star systems with a white dwarf
(WD) primary fed by a main-sequence donor via Roche-lobe overflow.
They represent a subclass of cataclysmic variables (CVs) and
experience semi-regular, luminous outbursts, during which accretion
onto the WD proceeds at a high rate. Most of the time, however, DN are
in quiescence --- a phase during which the accretion rate onto the WD
(and the system luminosity) is considerably reduced (see Warner 1995
for a review). According to the disk instability model (DIM), such a
behavior arises because most of the mass transferred by the companion
star builds up in an unsteady disk during quiescence (Cannizzo 1993;
Lasota 2001).

Quiescent DN are hard X-ray sources with typical luminosities of $\sim
10^{30}-10^{32}$~erg~s$^{-1}$  (see, e.g., C\'ordova \& Mason 1983;
Patterson \& Raymond 1985).  Spectral fits to this X-ray emission
suggest a Bremsstrahlung origin from gas with temperatures $\sim
2-20$~keV (Patterson \& Raymond 1985; Eracleous et al. 1991; 
Belloni et al. 1991; Yoshida et al. 1992; Mukai \& Shiokawa 1993). 
This X-ray emission is commonly attributed to the boundary layer (BL) 
at the interface between the WD and the thin accretion disk around it.
At low accretion rates ($\lsim 10^{16}$~g~s$^{-1}$, typical of 
quiescent DN), the gas in the BL is hot and optically thin, hence it 
is a substantial source of hard X-ray emission 
(Pringle \& Savonije 1979; Tylenda 1981; Patterson \& Raymond 1985). 
Detailed calculations by Narayan \& Popham (1993) show that
the optically-thin BLs of disk-accreting WDs can also be radially 
extended (of order a WD radius) and that energy
advection is an important element of their internal structure.

On the other hand, Meyer \& Meyer-Hofmeister (1994) presented two
arguments in favor of a more extended, inner hot flow structure in
quiescent DN. They noted that: (i) the standard DIM predicts accretion
rates at the WD surface which are smaller than those inferred from
quiescent X-ray luminosities by typically more than one order of
magnitude and (ii) the presence of an extended, low-density hot flow
in the WD vicinity may explain the observed delay (of $\sim 0.5-1$
day) in the rise to outburst of the EUV light relative to the optical
light (e.g., Mauche, Mattei \& Bateson 2001, but see Smak 1998).  Our
current understanding of the structure and properties of quiescent
disks is rather limited (Menou 2002), so studying the possible
presence of an extended, hot flow in quiescent DN is certainly
worthwhile.

Recently, Medvedev \& Narayan (2001a) discovered solutions for hot
accretion onto unmagnetized, rotating compact stars. They found that
at large stellar spin rates, dissipation in the hot flow is dominated
by stellar braking in a ``hot settling flow'' (HSF) configuration,
whereas in the opposite limit, the flow reduces to a conventional
advection-dominated accretion flow (ADAF; Narayan \& Yi 1994;
1995). Both flows exist at relatively low accretion rates
($\la10^{-2}$ of Eddington).  Medvedev \& Narayan (2001b) have also
addressed the thermal stability of their (cooling-dominated) settling
solution, and found that it is most likely stable, once {
turbulent} thermal conduction is accounted for.  The work of Medvedev
\& Narayan was largely focused on accretion onto neutron stars, but
the HSF solutions are also valid for WD accretion.  This is important
because, in general, much more is known about WDs in CVs than about
neutron stars in close binary systems. In particular, the rotation
rates of several WDs in DN have been measured during the last few
years thanks to {\it HST} spectroscopy (see Sion 1999 for a
review). In this {\it Letter}, we apply the model of Medvedev \&
Narayan to WDs in quiescent DN by constructing numerical solutions for
hot accretion in five systems with relatively well known system
parameters.

\section{System Parameters}

We have gathered existing data on quiescent DN for which estimates of
the WD rotation rate and the quiescent X-ray luminosity were both
available.  A summary of the system parameters adopted for this study
is given in Table~\ref{tab:one}. For the WD masses, $M_{\rm wd}$, and
orbital inclinations, $i$, we uniformly adopt the values given in the
catalog of Ritter \& Kolb (1998). To derive the WD rotation rates, we
use the observationally inferred values of $V_{\rm rot} \sin i$
compiled by Sion (1999) and make the assumption that the stellar
rotation axis is aligned with the rotation axis of the orbital motion.
A spin parameter, $s$, can then be calculated as the ratio of $V_{\rm
rot}$ to the Keplerian rotation rate at the WD radius, $V_{\rm
K}(R_{\rm wd})$. For the WD radii, we simply use the reasonable
mass--radius relation: $R_{\rm wd}=5 \times 10^{8}~{\rm cm}~ (M_{\rm
wd}/1.2~\msun)^{-0.33}$. Finally, references from which quiescent
X-ray luminosities were collected can be found in
Table~\ref{tab:one}. When several values existed for a given system,
we generally adopted the value corresponding to the hardest X-ray
spectral range (given the typically hard Bremsstrahlung spectrum of
quiescent DN). Note that the accuracy of the system parameters adopted
is not crucial because we only wish to illustrate the general
properties of hot accretion in these systems.\footnote{For instance,
even if the WD in WZ Sge was much more massive than assumed in
Table~\ref{tab:one} (e.g.  Steeghs et al. 2001), it would still be the
fastest rotator among all five systems, {as measured by the
$s$~parameter.}}

\section{Hot Accretion Solutions}

To investigate the nature of hot accretion in quiescent DN, we use the
numerical code described in detail by Medvedev \& Narayan (2001a).  We
solve the height-integrated, axisymmetric hydrodynamical equations for
the conservation of mass, radial and angular momenta, and two energy
equations for the proton and electron fluids. These equations are
solved by the relaxation method with appropriate boundary
conditions. A special grid was used to carefully resolve steep
gradients near the stellar surface. At the outer boundary, the
electron and proton temperatures and the angular velocity of the flow
are fixed to the values appropriate for an ADAF. At the inner
boundary, where the flow meets the stellar surface, the flow
temperature and the angular velocity are forced to match those of the
star. We assume a stellar temperature of $\sim{\rm few}\times10^5~{\rm
K}$ (the exact value is unimportant here). The mass accretion rate,
$\dot M$, is a free parameter, and a standard Shakura-Sunyaev
$\alpha$-viscosity is used. We assume that most of the viscously
generated heat goes to protons. However, this is unimportant because
the electrons are coupled to the protons via Coulomb collisions very
efficiently. The electrons are cooled via optically thin
Bremsstrahlung emission.  Thermal conduction and (self-absorbed)
synchrotron cooling are not included, for simplicity (the latter
should be negligible). The effects of Comptonization of the stellar
radiation and self-Comptonization of the Bremsstrahlung radiation are
included, but both are very weak. We generally adopt an adiabatic
index $\gamma=1.6$, close to the $\gamma=5/3$ of a non-relativistic,
monoatomic ideal gas.  In all models we choose the outer radius of the
flow to be at $\sim{\rm few}\times10^{11}~{\rm cm}$ to better isolate
the effects of the inner and outer boundary conditions. In real
systems, the outer radius is probably much smaller. The adopted
parameters, namely the stellar mass, $M_{\rm wd}$, its radius, $R_{\rm
wd}$, and its spin parameter, $s$, are given in
Table~\ref{tab:one}. In all systems but WZ~Sge, we used a conventional
value of the viscosity parameter, $\alpha=0.1$. In WZ~Sge we set
$\alpha=0.02$, as discussed below.

\begin{figure}
\plotone{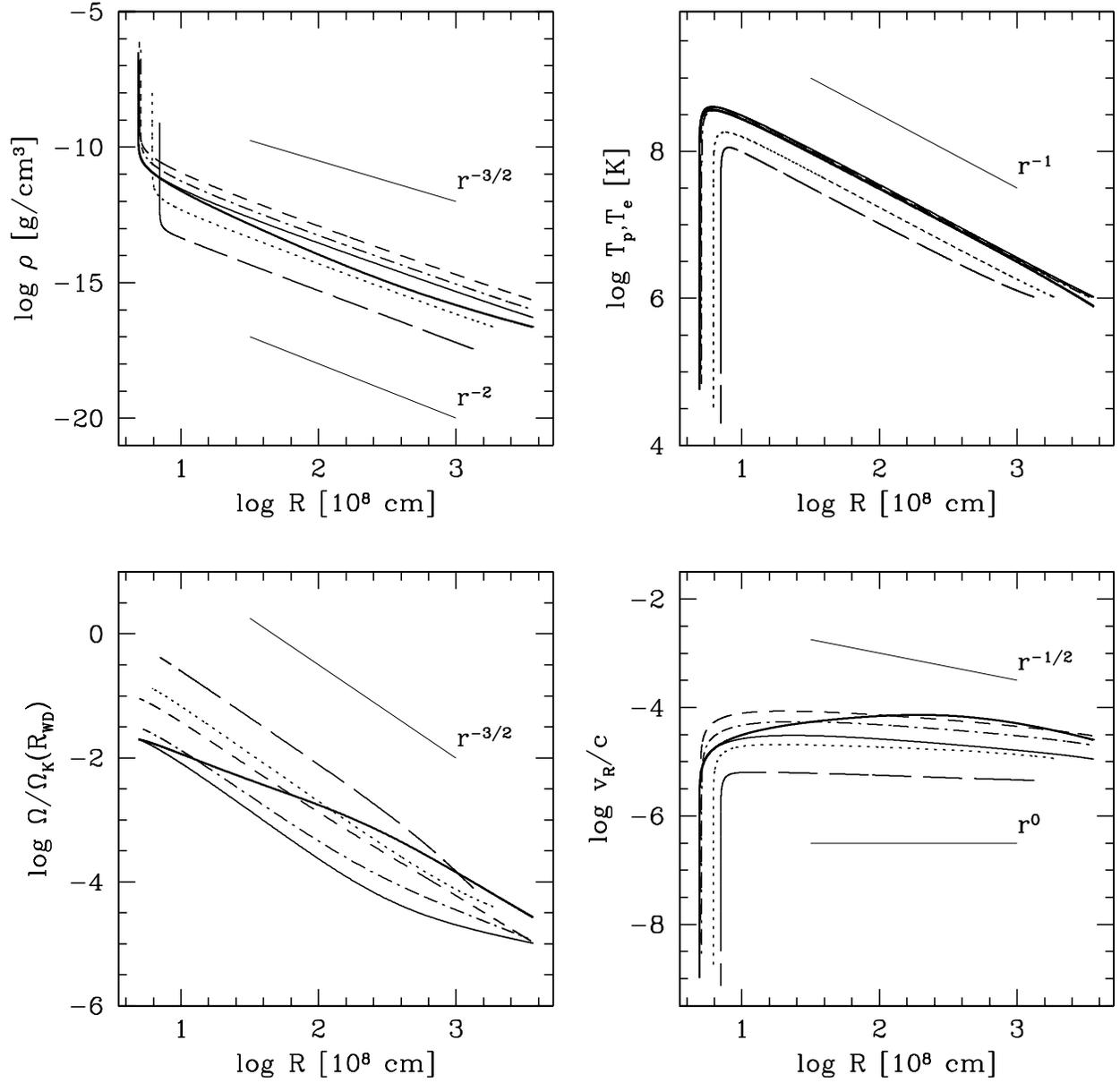}
\caption{Radial profiles of density, $\rho$, proton and electron
temperatures, $T_p$ and $T_e$, angular rotation velocity, $\Omega$,
and radial velocity, $V_R$, for hot accretion with $\gamma=1.6$ in the
five systems listed in Table~1 (RX And: dot-dashed, SS Cyg:
short-dashed, U Gem: thin solid, VW Hyi: dotted, WZ Sge: long-dashed).
The density profiles for VW Hyi and WZ Sge have been scaled down by a
factor 10 and 100, respectively. The thick solid line correspond to
the same model for U Gem except that $\gamma=4/3$ instead of $1.6$.
\label{fig:one}}
\end{figure}

Figure~\ref{fig:one} shows the numerical solutions obtained for the
five systems listed in Table~\ref{tab:one}. In each case, the
accretion rate in the steady, hot flow has been adjusted so that the
total Bremsstrahlung emission fits the quiescent X-ray luminosity 
given in Table~\ref{tab:one}. The four panels in Figure~\ref{fig:one} 
show the radial profiles of density, $\rho$ (in g~cm$^{-3}$), electron 
and proton temperatures, $T_e$ and $T_p$ respectively (in Kelvin; the 
curves overlap because $T_e\approx T_p$ indicating efficient Coulomb 
transfer of energy between the ions and the electrons), angular 
rotation velocity, $\Omega$ (in units of the Keplerian value at the 
WD radius, $\Omega_K(R_{\rm wd})$) and radial velocity,
$V_R$ (in units of the speed of light, $c$). 
The profiles are approximate power laws, with narrow boundary layers
close to the WD surface (except for $\Omega$, which smoothly matches
the stellar rotation rate). The profiles of $\Omega$ and $V_R$ 
indicate that boundary effects are present over a substantial range 
of radii, at both the inner and outer edges.

The scalings in an ADAF are $\rho \propto R^{-3/2}$ and $V_R \propto
R^{-1/2}$ (Narayan \& Yi 1994), while in a HSF they are $\rho \propto
R^{-2}$ and $V_R \propto R^{0}$ (Medvedev \& Narayan 2001a). The
profiles of $\rho$ and $V_R$ in Fig.~\ref{fig:one} deviate from
perfect power laws and are intermediate between the above two 
solutions. The Bernoulli number is negative everywhere (for 
$\gamma\sim5/3$), except in a narrow region at the outer edge of the 
flow where the ADAF conditions hold (in an ADAF, $Be>0$). Hence, the 
solutions should be convectively stable (Narayan et al. 2000; 
Medvedev \& Narayan 2001a).

The solutions shown in Fig.~\ref{fig:one} are not
``advection-dominated'' in the sense that the gas entropy decreases
inward (it increases inward in an ADAF).  The flow energetics is best
described by computing the ratio of the integrated Bremsstrahlung
luminosity to the available accretion luminosity, $L_{\rm acc}=GM_{\rm
wd} \dot M / R_{\rm wd}$. The accretion rates for the five systems are
quite low (Table~\ref{tab:one}). The corresponding ratios, $\eta_{\rm
eff} \equiv L_X/L_{\rm acc}$, are $\sim1$ for low-$s$ systems and
larger than unity for large $s$, indicating significant extraction of
rotational energy from the central star. Thus, in large-$s$ systems
such as WZ~Sge, the $\dot M$ value given in Table~\ref{tab:one} should
be considered as a rough upper limit because $L_X$ is essentially
powered by stellar spindown with little contribution from the
accretion luminosity. However, the luminosity of a HSF is sensitive to
$\alpha$. We were able to fit the observed $L_X$ of WZ Sge only for
$\alpha \approx 0.02$. Perhaps, this may indicate that the value of
$\alpha$ may be lower than generally expected.

The value of the adiabatic index has a strong effect on the flow
structure. We illustrate this by calculating a model for U Gem with
$\gamma = 4/3$ (all other parameters being the same). It is shown in
Fig.~\ref{fig:one} as a thick solid line. At large radii, the flow
structure now adopts an ADAF configuration, with $\rho \propto
R^{-3/2}$, $V_R \propto R^{-1/2}$ and a larger, constant value of
$\Omega / \Omega_K$. The Bernoulli number is positive everywhere in
the hot flow (excluding the BL) and energy advection is dominant in
the outer regions of the flow with ADAF properties.  However all the
advected energy is now radiated in the optically thin BL so that
$\eta_{\rm eff} \approx 1$ as well. This influence of $\gamma$ is
consistent with the results of Medvedev \& Narayan (2001a). According
to their classification, the solutions we obtained for $\gamma=1.6$
and the inner zone of the flow in the case with $\gamma=4/3$ are
``settling ADAFs'' {(WZ Sge clearly corresponds to a HSF,
however).}

\begin{figure}
\plotone{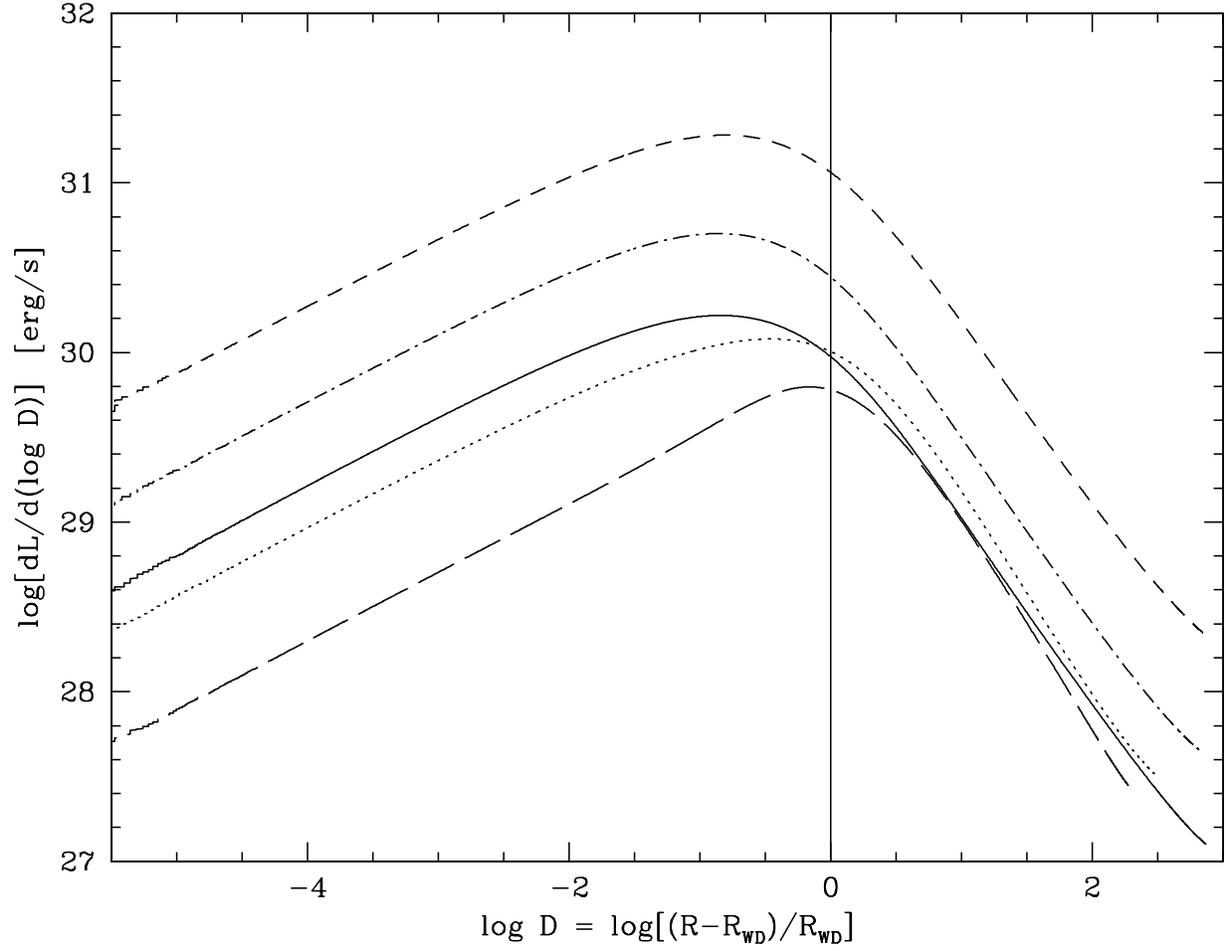}
\caption{Bremsstrahlung luminosity of the hot flow, per unit
logarithmic fractional distance from the white dwarf surface, for the
five models with $\gamma=1.6$ shown in Fig~\ref{fig:one} (same
notation). Most of the X-ray emission originates from the white dwarf
vicinity.
\label{fig:two}}
\end{figure}

Fig.~\ref{fig:two} shows the Bremsstrahlung luminosity, per unit
logarithmic fractional distance from the WD surface, for the five
solutions with $\gamma=1.6$ shown in Fig.~\ref{fig:one}. The density
in the flow is low enough that Comptonization has a negligible effect
on the emitted spectrum, even in the densest regions, near the WD
surface.  Fig.~\ref{fig:two} clearly shows that X-ray emission
originates mostly from regions in the close vicinity of the WD (the
same is true for the U Gem model with $\gamma=4/3$). The brightest
regions of the flow are also approximately the hottest, with
temperatures $\simeq 10^8$~K for WZ Sge, $\simeq 2 \times 10^8$~K for
VW Hyi and $\simeq 4 \times 10^8$~K for the three other systems.

\section{Discussion}

We constructed numerical
models for hot accretion onto unmagnetized, rotating WDs in five
quiescent DN. Our solutions are a significant improvement over
previous work on hot accretion in this context (Katz 1977; Kylafis \&
Lamb 1982; Mahasena \& Osaki 1999; Menou 2000) in that both the WD
rotation and the viscous nature of the flow are accounted for in the
present case.  This makes these solutions plausible modes of accretion
in quiescent DN.

The outer boundary conditions chosen are somewhat artificial and
constitute a significant shortcoming of the present study.  It is
unclear exactly how the transition from an inner hot flow to an outer
thin disk (known to be present) would occur. Strong boundary effects
may be expected for a hot flow with a limited radial extent.
Nonetheless, our work has the virtue of isolating the main properties
of the hot flow, independently of the disk.  It will be important
in the future to study the structure of a hot flow with a 
smaller radial extent and more realistic boundary conditions.  For the
non-relativistic gas considered here and non-dominant magnetic fields,
we expect rather large values of $\gamma$ and therefore flows which
are not subject to outflows (Blandford \& Begelman 1999) or convection
(Narayan et al. 2000; Quataert \& Gruzinov 2000).

Menou (2000) proposed that energy advection in a hot flow could be
important in powering a dominant EUV emission component that could
explain the strong He~II emission lines observed in many quiescent
DN. Since the hot flow solutions presented here lack energy advection 
(for $\gamma$'s near 5/3), this possibility seems unlikely.

While a star is spun up by accretion via a disk boundary layer (except
when rotating near breakup; Popham \& Narayan 1991; Paczynski 1991),
it is spun down when accreting via a hot flow like an ADAF, a HSF or
the hot flow solutions presented here (except at very low spin rates;
Medvedev \& Narayan 2001a). This property has potentially important
consequences for the spin history of WDs in DN. Sion (1999) emphasized
that in the standard picture WDs should rotate near break up in DN,
but observations suggest that rapidly spinning WDs are rare.  If a WD
is spun down via hot accretion during quiescence, an equilibrium at
substantially sub-Keplerian rotation rates may be expected from the
balance between spin-up in outburst and spin-down in quiescence. 
This possibility requires further study with more appropriate boundary
conditions.  We note that other mechanisms could contribute to
spinning down single WDs (Spruit 1998) and WDs in DN (and thus get rid 
of the angular momentum acquired during accretion), including angular 
momentum losses induced by coupling to an expanding envelope during 
nova explosions (Livio \& Pringle 1998; Sion et al. 2001).

X-ray eclipses in several quiescent DN indicate that the emission
originates from the vicinity of the WD (Wood et al. 1995; Mukai et
al. 1997; Pratt et al. 1999; Ramsay et al. 2001). Without a detailed
comparison, it is unclear whether the hot flow solutions presented
here satisfy the existing observational constraints or not. Such a
comparison should take into account the spectral coverage of the
various X-ray instruments used and it probably requires a reliable
boundary layer model  taking into account the role of heat
conduction in potentially modifying its emission properties (which is
beyond the scope of the present paper). It is certainly encouraging
that X-ray emission is highly concentrated near the stellar surface in
our models, but it is presently unclear whether these models will
be able to account for the available X-ray eclipse data or not. We
note, for instance, that the X-ray eclipse of OY Car observed by
Ramsay et al. (2001) suggest an X-ray flux approaching zero at eclipse
minimum and a small vertical extent for the X-ray emitting
region. These properties may not be easily reconciled with the hot
flow models presented here. In the future, it will also be possible
to probe the structure of hot flows in (even non-eclipsing) quiescent
DN with detailed X-ray spectroscopic diagnostics, as illustrated in
Menou, Perna \& Raymond (2001; see also Narayan \& Raymond 1999).

\section*{Acknowledgments}

We thank J. Raymond for comments on the manuscript.  Support for this
work was provided by CITA (for MM), and by NASA (for KM) through
Chandra Fellowship grant PF9-10006 awarded by the Smithsonian
Astrophysical Observatory for NASA under contract NAS8-39073. KM
thanks the Center for Astrophysical Sciences at Johns Hopkins
University for hospitality.

\clearpage

\begin{table*}
\caption{SYSTEM PARAMETERS}
\begin{center}
\begin{tabular}{ccccccccc} \hline \hline
\\
System & $M_{\rm wd}$ & $R_{\rm wd}$ & $i$ & $V_{\rm rot}$ & $s$ & $L_X$ & 
$\dot M$ & $\eta_{\rm eff}$ \\
name & ($\msun$) & ($10^8$~cm) & ($^\circ$) & (km~s$^{-1}$)& 
($V_{\rm rot}/V_{\rm K,wd}$) & (erg~s$^{-1}$) & (g~s$^{-1}$) & 
($L_{X}/L_{\rm acc}$) \\
\\
\hline
\\
RX And & $1.14$ & 5.1 & 51$^\circ$ & $190$ & $0.03$ & $3 \times 10^{31}$~$^a$ &
$10^{14}$ & 1.0\\
SS Cyg & $1.19$ & 5.0 & 37$^\circ$ & $500$ & $0.09$ & $10^{32}$~$^b$ &
$3.3\times10^{14}$ & 1.0\\
U Gem  & $1.26$ & 4.9 & 70$^\circ$ & $\leq 110$& $\leq 0.02$ & $10^{31}$~$^c$ &
$3\times10^{13}$ & 1.0\\
VW Hyi & $0.63$ & 6.2 & 60$^\circ$ & $460$ & $0.13$ & $7 \times 10^{30}$~$^d$ &
$3.5\times10^{13}$ & 1.5\\
WZ Sge & $0.45$ & 7.0 & 75$^\circ$ & $1240$ & $0.42$ & $3 \times 10^{30}$~$^e$ &
$1.1\times10^{13}$ & 3.2\\
\\
\hline
\end{tabular}
\label{tab:one}
\end{center}
NOTE. -- (a) Eracleous, Halpern \& Patterson (1991) (b) Yoshida, Inoue
\& Osaki (1992); Mukai \& Shiokawa (1993) (c) Szkody et al. (1996) (d)
Mukai \& Shiokawa (1993); Eracleous et al. (1991) (e) Eracleous et
al. (1991); Mukai \& Shiokawa (1993).
\end{table*}

\end{document}